\documentclass{aa}  

\usepackage{graphicx}
\usepackage{txfonts}
\begin{document}

   \title{Mass flows in the Galactic Center by supernovae of the circumnuclear disk}

   \titlerunning{Supernovae of the circumnuclear disk}
   %\subtitle{I. Overviewing the $\kappa$-mechanism}

   \author{B. Barna,
          \inst{1}
          R. Wünsch,\inst{2}
          J. Palous,\inst{2}
          M. R. Morris,\inst{3}
          S. Ehlerová,\inst{2}
          P. Vermot\inst{4}
          %\and
          %C. Ptolemy\inst{2}\fnmsep%\thanks{Just to show the usage
          %of the elements in the author field}
          }

   \institute{Physics Institute, University of Szeged, D\'{o}m t\'{e}r 9, Szeged, 6723, Hungary\\
              \email{bbarna@titan.physx.u-szeged.hu}
         \and
             Astronomical Institute, Academy of Sciences, Bo\v{c}n\'{\i} II 1401, Prague, Czech Republic
             %\email{c.ptolemy@hipparch.uheaven.space}
        \and
            Department of Physics and Astronomy, University of California, Los Angeles, CA 90095-1547, USA
        \and
            LESIA, Observatoire de Paris, Université PSL, CNRS, Sorbonne Université, Université de Paris Cité, 5 place Jules Janssen, 92190 Meudon, France
             }

   \date{Received November 25, 2024; accepted XX, YY}

% \abstract{}{}{}{}{} 
% 5 {} token are mandatory
 
  \abstract
   {The circumnuclear disk  (CND) is presently the main supply of mass for the accretion onto the supermassive black hole (SMBH) in the Galactic Center (GC). While the accretion is relatively slow, it has been suspected that local episodic explosive events play an important role in the temporary mass inflow toward the SMBH, while also affecting the evolution of the CND.} 
   {The aim of this study is to follow the changes in mass flows caused by  supernova (SN) explosions nestled in or near the CND.}
   {We perform simulations with the grid-based MHD code FLASH of the inner 5 pc of the Milky Way GC, including gravitational potential, rotation, magnetic field, central wind source, and the warm gas of the CND, all mimicking the observed physical properties.}
   {Assuming a M$_\mathrm{SN}=10$ M$_\odot$ as the mass of the precursor of the core-collapse SN event at various locations within 2 pc from the GC, we detect a temporary increase in the accretion rate, transferring an additional 2-60 M$_\odot$ of warm gas to the immediate vicinity of the SMBH, depending on the explosion site. At the same time, the kinetic energy of the SN  blows away even  mass from the CND; the additional warm gas leaving the simulation domain after the explosion is on the order of $\sim100$ M$_\odot$. In the studied cases, the impact on mass flows and the turbulence caused by the explosion cease after $\sim250$ kyr.}
   {}


   \maketitle
%
%-------------------------------------------------------------------

\section{Introduction}

   The innermost few parsecs of the Galactic Center (GC) are occupied by a structure formed by dense molecular gas, called the Circumnuclear Disk (CND). Our knowledge about the properties of the CND has been considerably enhanced by observations with ALMA and SOFIA. The size and form of the CND (or, the Circumnuclear Ring, CNR, which is the inner edge of the disk) vary in the literature, with a typical inner radius of 1.5 - 2 pc. The outer radius is less well-defined because of the lower densities with continuous transition into the rarified environment. Here, the motion of the molecular gas within 1 pc is dominated by the gravitational potential of the SMBH and further away it follows a Keplerian rotation in the gravitational field of the SMBH and of the Nuclear Star Cluster (NSC). 

With an estimated mass of a few tens of thousands M$_\odot$, the CND has a significant impact on the mass transfers and structural evolution near the GC. Observations show that the disk surrounds the so-called mini-spiral consisting of gas streams that transport relatively dense gas to the close vicinity of Sgr A*, and consequently contribute to the growth of the SMBH. The disk itself is not homogeneous, but is populated with clumps and filaments. 
The group of local massive stars, also called the young nuclear cluster (YNC), nestled in the NSC within 0.5 pc of Sgr~A*, issues strong stellar winds that exert a powerful impact on the local environment. This wind ablates or drags the low-density outer parts of the CND and carries mass to high Galactic latitudes, producing outflows from the galactic disk. These processes are enhanced by the rapid outburst of kinetic energy caused by supernovae (SNe).

The formation mechanism and long-term evolution of the CND have not been understood yet. Some 
%of the theories \citep[e.g.][]{Güsten87,RequenaTorres12} about its origin assume 
argue that the CND is a transient structure \citep[e.g.][]{Güsten87,RequenaTorres12}. Such hypotheses generally claim that an infalling gas cloud encountered the SMBH and was circularized by tidal forces that left it in its configuration as a disk \citep{Sanders98,
Bradford05, Bonnell08, Wardle08, Hobbs09, Alig11, Oka11, Mapelli12, Mapelli16,
Goicoechea18, Trani18, Ballone19}. The non-uniformity is supposed to be smoothed out by differential rotation in a couple of dynamical timescales (i.e. a few hundred thousand years). Similarly, \cite{Güsten87} showed that the observed dense clumps in the disk are tidally unstable with a lifetime of $\sim$10$^{5}$ yr, 
%thus, it is likely 
so they argued that they would exist only in a young, transient disk. On the other hand, it was pointed out that clumps might only be artifacts caused by observational limitations \citep{Dinh21}. Violent episodic events, like SNe arising from inside or near the CND, could create transient structures. The estimated SN rate for the far larger and more massive Central Molecular Zone ($M_\mathrm{CMZ} = 3 - 6 \times 10^{7}$ M$_\odot$) is $\sim 2 - 5 \times 10^{-4}$ yr$^{-1}$ \citep{Dahmen+98, Crocker11}. At the same time, the SN rate for the three orders of magnitude less massive CND is not well constrained. However, proportionally more SN remnants can occur in the vicinity of the CND because of the strongly peaked stellar density and the estimated top-heavy initial mass function in the YNC \citep{LuJR+13}. %and because of the presence of the YNC hosting at least 100 massive stars, which may occasionally scatter out to the disk, before undergoing supernova explosions. 
The high number of SN Ia precursors inferred from X-ray surveys indicates that such SN explosions are also common there.
Considering both of these contributions, we can assume that the SN rate should be particularly high in the vicinity of the GC. We estimate that the average time between SNe occurring in or near the CND is comparable with, or even shorter than, the dynamical timescale of the disk ($t_\mathrm{dyn} \simeq 100$ kyr), thus providing a continuous impact on its evolution by stellar explosions.

The evolution and longevity of the CND have been subjects of numerous hydrodynamic modeling efforts. \cite{Coker03} used the finite-difference Eulerian ZEUS3D code \citep{Clarke96} to study the formation of the CND by simulating the infall of single giant  and multiple smaller molecular clouds. A similar simulation was performed by \cite{Mapelli16} by using the GASOLINE N-body smoothed particle hydrodynamic (SPH) code. \cite{Dinh21} performed a set of 3D SPH simulations, each including the external and the self-gravity, artificial viscosity, and varying turbulence. A greater volume, the innermost 20 x 20 x 10 pc of the GC, was modeled by \cite{Blank16} with the AMR MHD code AstroBEAR \citep{Cunningham09} through several dynamical times; their work showed the impact of magnetic field on the inward migration of mass. \cite{Solanki23} used Athena++ \citep{Stone20}, a grid-based magneto-hydrodynamic code, to simulate the interaction between the central wind and the inner edge of the CND, causing accretion to smaller radii. 

Despite the importance of high-energy shocks in the dense CND environment, individual SNe have not been included in similar hydrodynamic simulations. In this current study, we aim to follow the long-term evolution of an accreting disk representing the CND of the Milky Way GC. In this study, we improve the model by adding occasional SN explosions to the dynamical investigation of the CND. 
This extension is motivated by the presence of the YNC, which hosts core-collapse SNe at a higher rate than the dynamical timescale of the CND, and because SN Ia might be abundantly occurring in this high-stellar-density region. By following the evolution of the SN remnant, we investigate the impact of its kinetic energy on the mass flow towards the GC, thereby constraining the ultimate accretion onto the SMBH and the rate at which gas is ejected from the CND. 

This paper is organized as follows. In Section \ref{sec:cnd} we describe the observed physical properties of the CND, and how our model setup reproduces these features. In Section \ref{sec:results}, we show and interpret the evolution of our fiducial model when a supernova is added to the setup, and compare the model results to the observational features.  We summarize our conclusions in Section \ref{sec:conclusions}.

\section{Simulation of the Circumnuclear Disk}
\label{sec:cnd}

\subsection{Observed CND properties}
\label{sec:cnd_prop}
 The model disk, just like the gravitational potential and stellar wind source described in Sect. \ref{sec:model_cnd}, is assumed to mimic the mass distribution in the center of the Milky Way. Observations show a sharp inner edge of the disk at $r_\mathrm{GC}=1.5$ pc \citep{Christopher05} and a less distinct outer radius with  $r_\mathrm{GC}=3-7$ pc \citep{Hsieh21}. The thickness of the disk linearly increases from 0.4 \citep{Jackson93} to 2 pc \citep[depending on the exact location of the outer edge;][]{Vollmer01}, respectively. The temperature of the disk close to the rotational plane is in the range of 200 - 500 K \citep{RequenaTorres12, Mills+13}, which is too low to maintain the observed CND thickness alone via thermal pressure. 

Another important component is the magnetic field, whose strength is in the order of $B \simeq  1-10 \times 10^{-3}$ G as it was shown by observation of Zeeman splitting \citep{Plante95}, MHD shocks \citep{Bradford05} and the polarization of dust particles \citep[e.g.][]{Hildebrand93,Guerra23,Akshaya24}. The impact of the $B$ field can be described by the plasma beta parameter, which is the ratio of thermal-to-magnetic pressure components:

\begin{equation}
    \beta = \frac{n\,k\,T}{B^{2} / 2}
\end{equation}

where $n$ and $T$ are the number density and the temperature of the gas, respectively, while $k$ is the Boltzmann-factor. The observations provide a value of $\beta \simeq 0.001$, however, the kinetic pressure arising from the presence of strong turbulence probably exceeds the thermal pressure. Studies based on the observed velocity dispersion give estimates between 0.03 \citep{Mills17} and 0.7 \citep{Hsieh21} for $\beta'$, which also incorporates the impact of turbulence-associated pressure on the magnetic pressure. Either way, the modeling attempts of the CND cannot fully describe its evolution without the inclusion of magnetohydrodynamic (MHD) effects.

The observations also suggest that the disk is not uniform, instead, it shows a clumpy morphology \citep[e.g.][]{Vollmer01, Montero09, Hsieh21}. The fragments could be significant indicators of the age of the CND depending on whether these features are dense enough to be tidally stable, and thus, possibly exist on several dynamical timescales.
Other sub-structures within the CND are the 
%streams leading to the inner cavity with a radius of 1.0-1.5 pc. 
ionized gas streams located in the inner cavity  transporting matter closer to the GC (namely the northern and eastern streams, constituting two of the three arms forming the so-called Mini-Spiral). Due to the angular momentum loss of the gas flow, which is potentially enhanced by collisions of the streams and by self-collision and concomitant circularization, some matter is captured by the accretion disk of the Sgr A* and contributes to the growth of the SMBH \citep{Liszt03,Paumard04,Zhao09}. 
The inner cavity is partially maintained by the strong magnetic field. \cite{Blank16} found that the assumed milligauss azimuthal field stabilizes the inner disk rim against rapid inward migration of mass caused by the %cogwheel instability 
instabilities due to the interaction of the wind with the inner edge of the disk.

\begin{figure*}
    \centering
	\includegraphics[width=16cm]{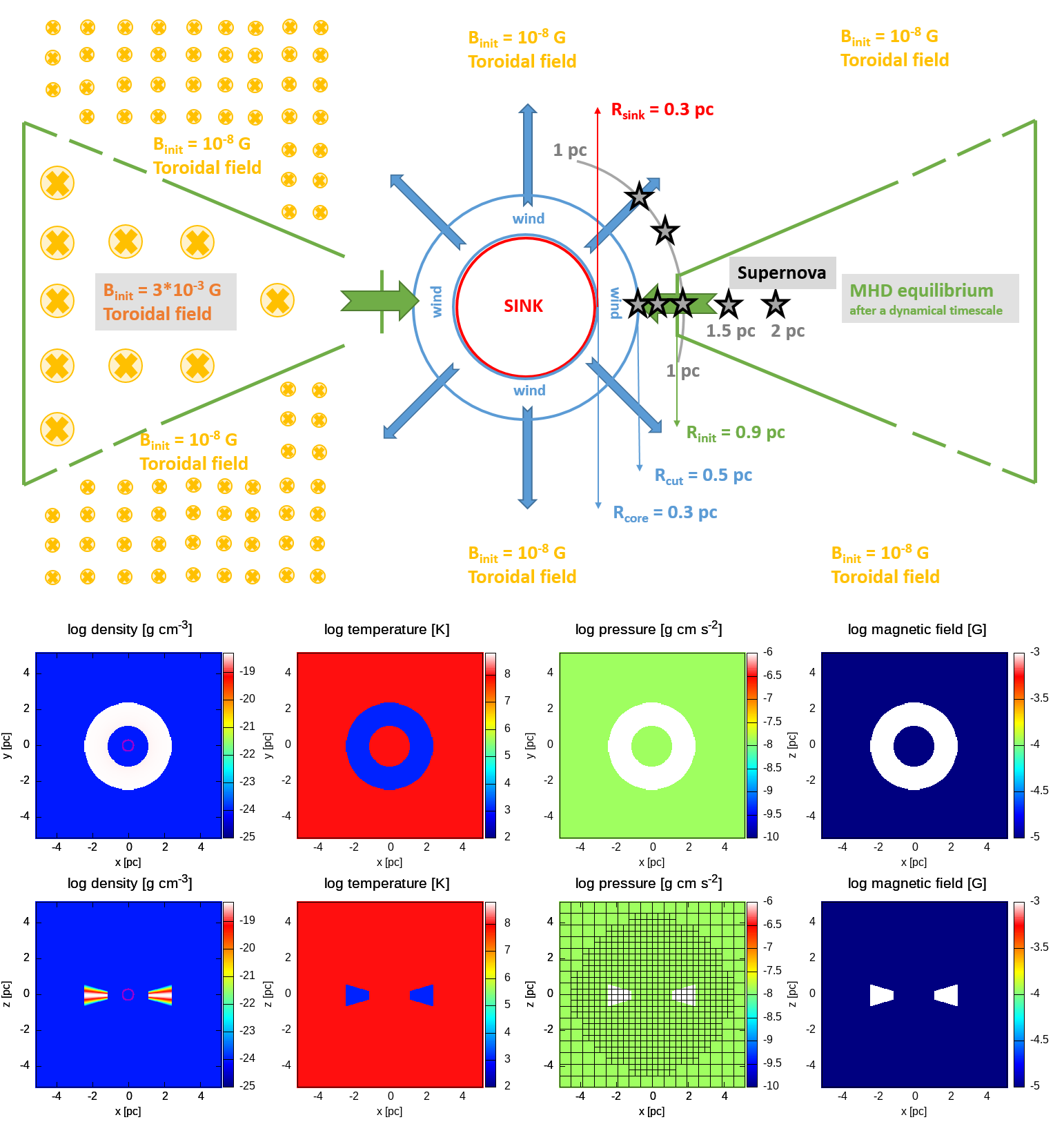}
    \caption{The initial properties of the CND in our model setup at T=0, with the disk shown in cross-section at the top, followed by face-on plots and cross-sectional plots of the initial values of physical parameters. The grid in the pressure plot (bottom line) represents the refinement of the simulation domain.
    }
    \label{fig:cnd_init}
\end{figure*}

\begin{figure}
    \centering
	\includegraphics[width=\columnwidth]{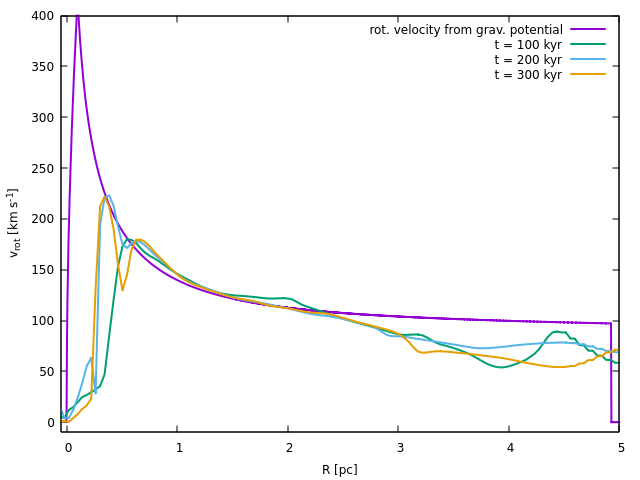}
    \caption{The evolution of rotational velocities in the rotational plane (at $z=0$).
    }
    \label{fig:rotvel}
\end{figure}

\subsection{Grid-based model of the CND}
\label{sec:model_cnd}

In this study, we use the 3D adaptive mesh-refinement code FLASH \citep{Fryxell00} for modeling the CND over several dynamical times, $t_\mathrm{dyn} \simeq 150$ kyr. Since the aim of this study is the long-term follow-up of CND evolution, the environment (i.e. gravitational potential, stellar wind, ambient medium properties) is designed to be stationary over multiple dynamical times of the CND. Note that most of the numerical descriptions handle the CND independently from the medium around the disk. Moreover, considering the inaccuracies inherent to grid-based simulation (e.g. numerical viscosity, insufficient resolution), the modeled CND slightly deviates from the theoretical descriptions. %This method will result that, after a phase of oscillation, the disk properties relax into a quasi-equilibrium state, similar to the original design.
As a result, the disk properties, will, after an initial oscillatory phase, relax into a quasi-equilibrium state, similar to the initial design.
% if the relaxed state is similar to the initial conditions, then why does it oscillate.  Maybe something more needs to be said besides "similar"?

The initial structure of the CND (the cross-section can be seen in Fig. \ref{fig:cnd_init}) follows hydrostatic equilibrium  governed by the ratio of the rotation speed of the disk to the sound speed (magnetic pressure is not accounted for yet). In the isothermal disk, if no other source of pressure is introduced, the gas pressure would provide a Gaussian profile for the initial vertical density distribution (i.e., along the z-axis), with a scale height of $H = c / \Omega$, where $c$ is the sound speed and $\Omega $ is the angular orbital speed in the CND. We assume a constant density $\rho_0$ in the x-y rotational plane ($z = 0$), between the inner ($R_\mathrm{min}$) and outer edge ($R_\mathrm{max}$) of the CND.

Initially, the torus covers the range from $R_\mathrm{min}=1.0$ to $R_\mathrm{max}=2.5$ pc with a density in the rotational plane of $\rho_{0} = 5*10^{-19}$ g cm$^{-3}$. The sound speed, and thus the scale height, is prescribed only by the $T_\mathrm{CND}$ temperature of the disk. During the simulation, the interior of the CND is dense enough to effectively cool, following the prescription of \cite{Schure09} and assuming solar metallicity.  However, in such a case, $T_\mathrm{CND}$ would drop below $\sim$100 K and our resolution would not allow us to properly simulate such high density regions. This would lead to the collapse of the disk into a thin layer, which does not correspond to observations. To avoid this, we introduce a floor value, which artificially prevents the temperature from dropping below $T_\mathrm{CND}=1000$ K. 
This is assumed to partially cover the effect of microturbulence, which is mainly responsible for the pressure support, but not included in our setup due to the insufficient resolution (see \citet{Solanki23} for a similar approach). Note that our setup also includes a magnetic field which also contributes to the (magneto-hydrodynamic, MHD) equilibrium of the disk via turbulence and magnetic pressure.

The mass of the initial setup amounts to $M_\mathrm{CND}$ = 22000 M$_\odot$, which matches with the estimated mass of the CND.  Initially, although the magnetic pressure is accounted for, the outer regions of the disk expand because the initial setup was not in MHD equilibrium (see below). As a result, the density drops to $\sim 10^{-20}$ g cm$^{-3}$ in the rotation plane, and the outer, low-density, thicker extensions of the disk (at a greater galactocentric distance, $r_\mathrm{GC}$ > 3 pc) start oscillating with decreasing amplitude. The total CND mass rapidly decreases mainly due to the wind, which sweeps away $\sim$1600 M$_\odot$ from the lower-density outer lobes of the disk during the first 100 kyr. After the relaxation of the simulation, the warm gas of the disk dragged by the wind integrates into the outflow resulting in a mass-loaded wind with higher density than the initially inserted wind source. At $t = 250$ kyr (the moment of the imposed SN explosion), the steady escape rate of mass from the computational domain is $\sim$0.45 M$_\odot / kyr $, which corresponds to a mass loading factor of 3 in the escaping wind.

\begin{figure*}
    \centering
	\includegraphics[width=16cm]{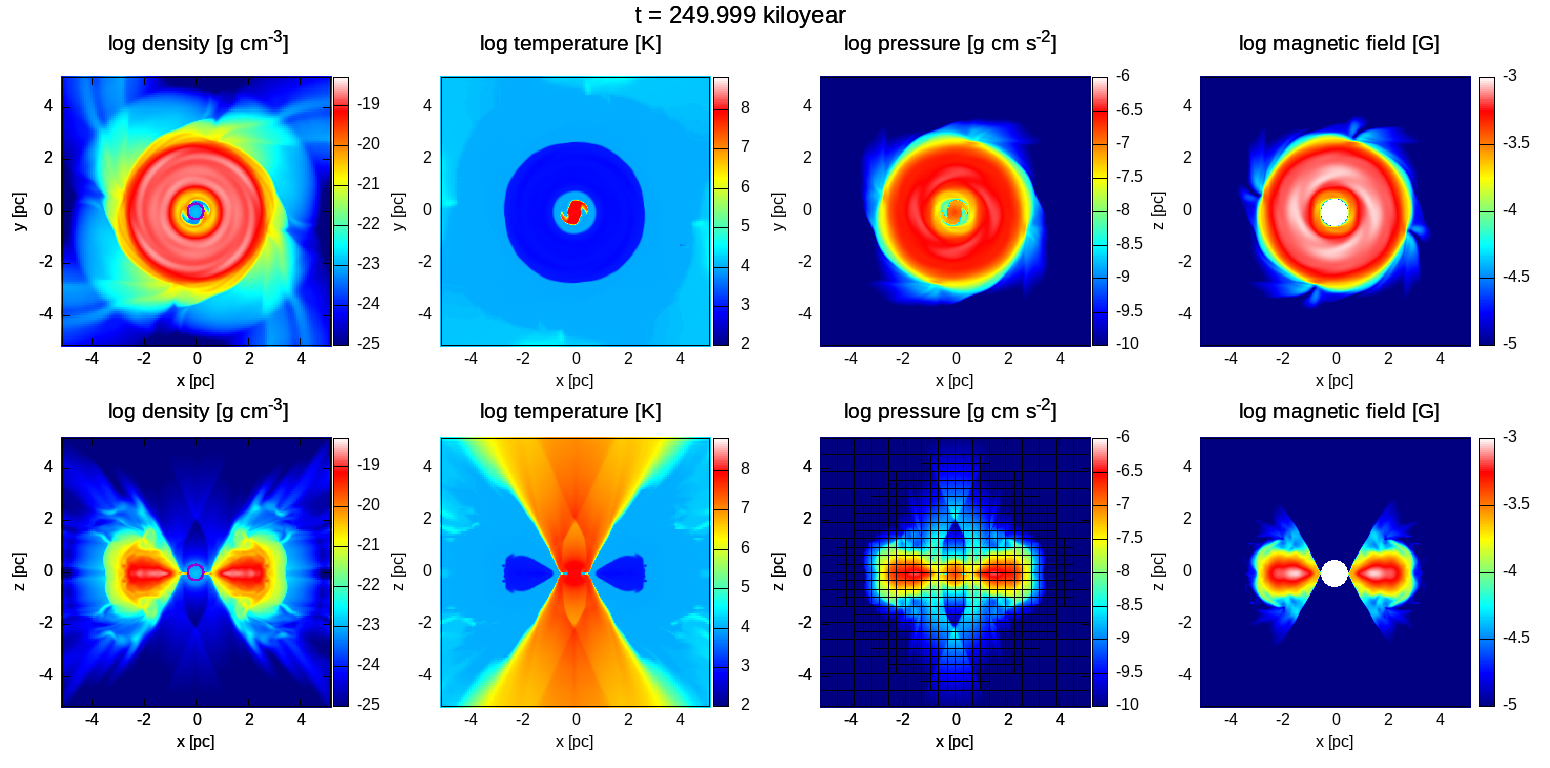}
    \caption{CND properties at $t = 250$ kyr in the fiducial simulation. Same panels as in Figure 1.}
    
    \label{fig:ref_sim1}
\end{figure*}

\begin{figure*}
    \centering
	\includegraphics[width=16cm]{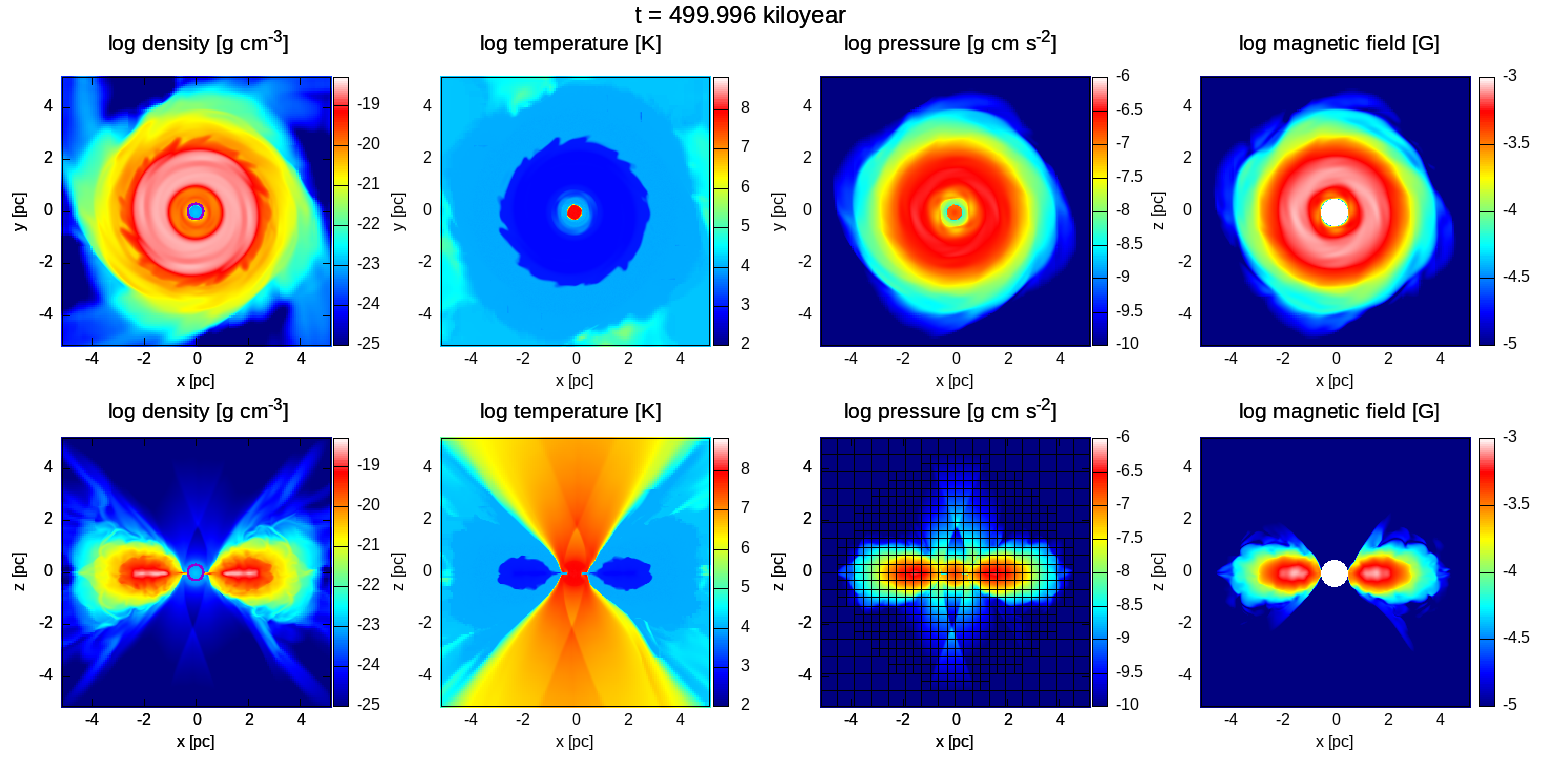}
    \caption{Relaxed CND properties after $t = 500$ kyr in the fiducial simulation.}
    
    \label{fig:ref_sim2}
\end{figure*}

\begin{figure*}
    \centering
	\includegraphics[width=16cm]{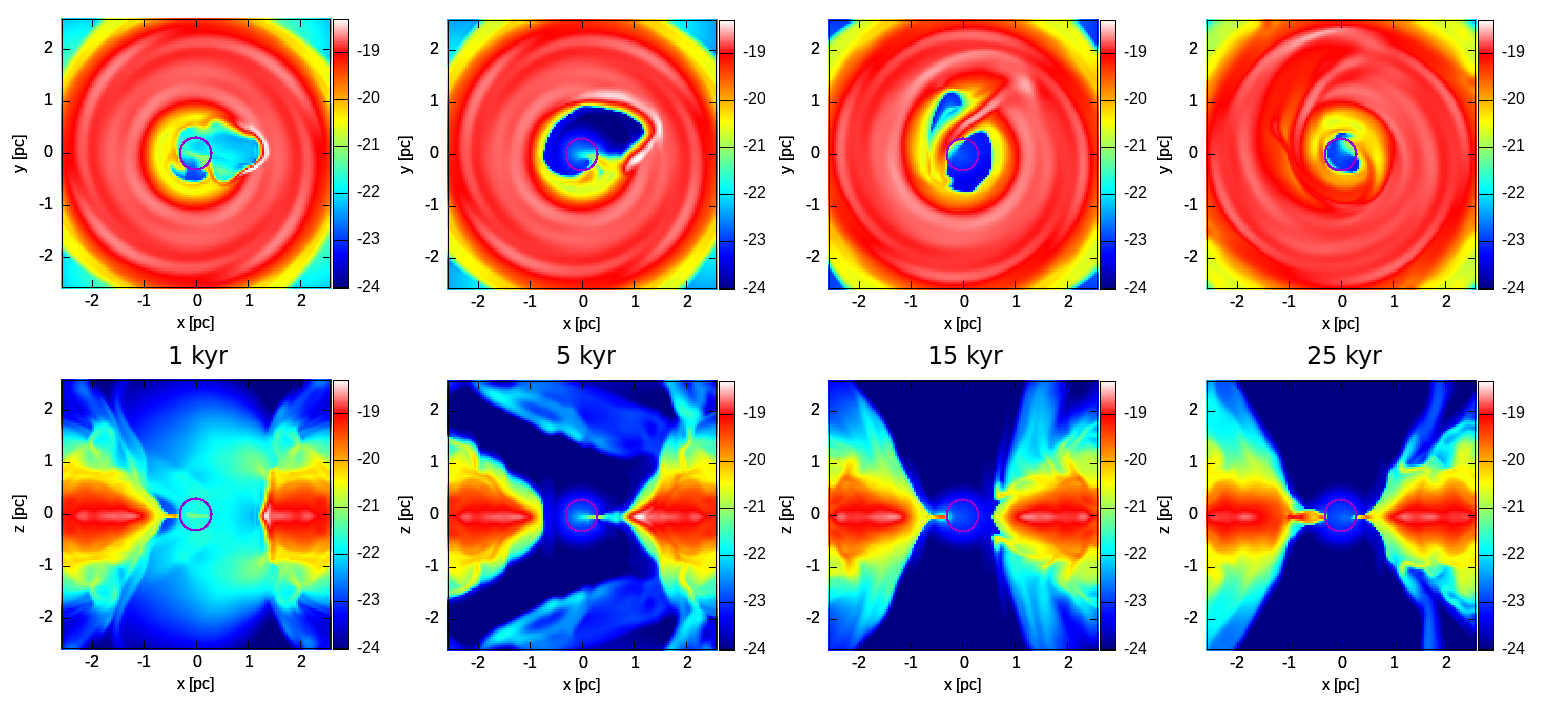}
    \caption{The evolution of the SN remnant with explosion site at $r_\mathrm{GC}=1$ pc in the midplane of the CND (model SN\_r10\_i00).}
    
    \label{fig:sn1_evolution}
\end{figure*}

The adopted gravitational potential follows the description of \cite{Chatzopoulos15}, assuming a point-like SMBH and two-component NSCs:

\begin{equation}
\Psi (r_\mathrm{GC}) = -\frac{G M_\mathrm{SMBH}}{r_\mathrm{GC}} - \sum_{i=1}^{2}\frac{G M_\mathrm{NSC_1}}{a_\mathrm{i}(2 - \gamma_\mathrm{i})} \Bigg(1 - \bigg(\frac{r_\mathrm{GC}}{r_\mathrm{GC} + a_\mathrm{i}}\bigg)^{2-\gamma_\mathrm{i}}\Bigg),
	\label{eq:potential}
\end{equation} 
\\

\noindent
where $a_\mathrm{1}$ = 3.9 pc, $\gamma_\mathrm{1}$ = 0.51 and $a_\mathrm{2}$ = 94.4 pc, $\gamma_\mathrm{2}$ = 0.07 are the characteristic sizes and the density slopes of the two components of the NSCs, respectively \citep[similarly to the previous hydrodynamic simulations of the GC, e.g.][]{Palous20, Barna22}. 

In a realistic disk, the rotational velocity deviates slightly from the function ($v_\mathrm{0}$) estimated from Eq. \ref{eq:potential} due to the radial pressure gradient. This difference is second order in $c / v_\mathrm{0}$, which, considering the disk temperature of $T_\mathrm{CND} = 1000$ K, is $\sim$2-3 km\,s$^{-1}$, depending on the radial coordinate. Since early oscillation of the disk properties, including the velocity profile, are higher due to the interaction with the ambient medium than the effect of the velocity adjustment, we neglect the deviation from the initial $v_\mathrm{0}$.

This initial rotational velocity field is valid only within $0.1 < r_\mathrm{GC} < 5$ pc. At the very center within $r_\mathrm{GC} = 0.1$ pc, the gravitational potential is set to be constant to avoid very high velocities. Beyond $r_\mathrm{GC} = 5$ pc, at the beginning of the simulation, a static outermost region is defined, where both the gravitational potential and the initial rotation velocity are set to zero. This method is required to avoid the formation of voids at the corners of the simulation volume due to the rotation. The \textit{diode} option, allowing only one-way outflows, is introduced as the boundary condition for the setup. As a result, the matter reaching $r_\mathrm{GC} = 5$ pc can leave the domain without any obstacles, thus preventing unpredictable inflow into the simulation domain in the X- and Y- directions. This treatment of the outermost region has no impact on the mass flows within the CND. Material that reaches this outer region from the wind source or the CND may flow out and leave the simulation domain.

The velocity cross-section of the simulation domain deviates from the initial function in the mid-plane and slows down with time due to the boundary interactions and the oscillating expansion of the disk. The deviations from Keplerian velocities appear mainly within $r_\mathrm{GC} < 0.5$ pc, where the sink captures the inward-moving mass with its momentum (see in Sect. \ref{sec:central}). Outside of $r_\mathrm{GC} > 3$ pc, the low-density extension of the disk are affected by the outward moving flows. At the same time, the great majority of the CND maintains the initial velocity profile, and its structure (c.f., Fig. \ref{fig:rotvel}). 

The resolution of the model is constrained by needing to reproduce the main components of the mass flows, but, at the same time, to keep the simulation runs manageable with computational resources of $\sim$1000 CPU. Therefore, the resolution of the 10x10x10 pc simulation domain is set according to the steady-state accretion rate of the CND model, which matches the estimated mass inflow rate of the GC (for a more detailed description  see Sect. \ref{sec:acc_rate}).
%The outer region of the domain (r$_\mathrm{GC} > 4$ pc), to which the dense layers of the disk do not extend, is resolved with one refinement level lower.

As an indicator for the stability of our model, we check that the disk keeps its integrity and maintains a Gaussian density profile along the z-axis, and that it deviates only slightly from axial symmetry at large $R$ (see in Figs. \ref{fig:ref_sim1} and \ref{fig:ref_sim2}). Additionally, we check that, the $v_{rot} (R)$ rotational curve keeps its initial function over multiple dynamical timescales outside the $r_\mathrm{GC} = 0.1$ pc zone (Fig. \ref{fig:rotvel}). Discrepancies with these conditions emerge only in the outermost regions ($R > 3$ pc) where the low-density outer extensions of the disk expand and radially oscillate due to the impact of stellar wind and the temporary lack of MHD equilibrium.

\subsection{Magnetic field}
\label{sec:mag_field}

%%%%
Polarization vectors indicate that the B field lines have a dominantly azimuthal orientation \citep{Werner88} on a large scale, since the magnetic field follows the differentially rotating gas around the GC \citep{Wardle90}. \cite{Guerra23} showed that magnetic pressure is orders of magnitude stronger than gas pressure, thus, former options are more likely. Here we assume a strong magnetic field as the main contributor maintaining the CND thickness perpendicular to the rotational plane, and it evolves until it reaches MHD equilibrium. 

At the beginning of the simulation, a purely azimuthal magnetic field is introduced. It follows the rotation of the gas and has an initial value of $B_\mathrm{init} = 3 * 10^{-3}$ G in the CND. For the environment, $B_\mathrm{init}$ is scaled with the density because of the strong coupling between matter and magnetic $B$ field in the MHD solver. The minor inconsistencies of the initial setup, such as the jump in the $B$ field at the surface of the CND, quickly vanish because of the movement of the matter, which is coupled to the magnetic field lines. Similarly to hydrodynamical properties, the magnetic field also relaxes with time (Figs. \ref{fig:ref_sim1} and \ref{fig:ref_sim2}). Weak $B_\mathrm{z}$ component (that which is perpendicular to the rotational plane) emerges following the expansion of the disk, but remains two orders of magnitude smaller than the other components. The central wind source is inserted without any magnetic field component on the grid, in accordance with the random orientation of stars of the NSC.

By the moment of the SN explosion, the simulaton has reached MHD equilibrium. In the rotational plane, $B$ varies between 0.0005 and 0.001 G, which matches well the polarization measurements of the Galactic Center. The field weakens outward with the decreasing density and it is negligible in the medium around the CND. The approximate value of $\beta$, which also accounts for the turbulent pressure in the CND (see in Sect. \ref{sec:model_cnd}), is in the range of 0.1 - 0.5, meaning that magnetic pressure plays an important role in the formation of structures in the disk. To capture this kind of evolution, relatively long simulations are needed, at least a few times the dynamical timescale ($t_\mathrm{dyn} \simeq 150$ kyr). In the simulation with magnetic field, stellar wind is confined by the magnetic pressure, since the local value of $\beta$ in the wind is an order of magnitude lower than that of the CND.

\subsection{Central region: sink and wind source}
\label{sec:central}

The accreted mass from the CND can reach the central 0.5 pc, where the complex structures and mass flows (namely ionized gas streams, WR winds, the accretion flow, orbiting compact clouds, outflow from the inner accretion flow) are not resolved in our simulations. Here, a large fraction of the gas pulled or driven into the gravitational well reaches the close vicinity of the SMBH from where it may join the accretion flow. This process is modeled by a central sink with a radius of $r_\mathrm{sink} = 0.3$ pc, corresponding to the region where the gravitational potential of the SMBH dominates over other processes considered in our MHD simulation. The simulation eliminates mass that moves within $r_\mathrm{sink}$, as well as its corresponding momentum, if its density surpasses the threshold value of $n_\mathrm{sink} =500$ cm$^{-3}$, assuming that that mass has inexorably joined the accretion flow. Note that the value $n_\mathrm{sink}$ affects only the time at which the mass accumulation at the center reaches a quasi-steady rate, and thus, it can be replaced by other number densities. The chosen threshold of $n_\mathrm{sink} =500$ cm$^{-3}$ provides measurable accumulation in the accretion disk within $\sim$50 kyr, which is significantly shorter period than $t_{\mathrm{dyn}}$.
%and also allows us to take into account the dense molecular phase of the ISM. %Note that the sink parameters are chosen to provide measurable accretion within a relatively short period (other values may delay or speed up the start of accretion), but the values are arbitrary and represent no physical meaning.

We also assume a spherically symmetric central wind originating from the young population of the NSC 
with the mean radius of 0.5 pc. 
We model the impact of the $\sim$100 massive stars known to populate the YNC \citep{LuJR18}. The same treatment is adopted as that in the study of \cite{Ehlerova22}, assuming a wind luminosity $L = 3 \times 10^{38}$ erg\,s$^{-1}$ and a wind terminal velocity of $v_{\inf} = 2500$ km\,s$^{-1}$. These provide a mass injection rate of $\dot{M}_w = 2\,L / v_{\inf}^{2} = 1.5 \times 10^{-4}~ $M$_\odot$\,yr$^{-1}$ as the basis of the central wind. %regardless of the mass loading. 

The density of the wind source, n$_\mathrm{w}$, is a function of the radial coordinate r$_\mathrm{GC}$ in the form:

\begin{equation}
n_\mathrm{w}(r_{\mathrm{GC}}) = \frac{n_{\mathrm{w0}}}{\left[1+\left(\frac{r_{\mathrm{GC}}}{r_{\mathrm{w0}}}\right)^2\right]^\beta},
\label{eq:wind}
\end{equation}

\noindent
where volume density $n_\mathrm{w}(r_{\mathrm{GC}})$ is specified with the following fitting parameters: $n_{0, \mathrm{w}} = 1\, \mbox{cm}^{-3}$, $\beta = 1.5$, $r_{\mathrm{w0}} = 0.3\ \mbox{pc}$. 

Some fraction of the inflow from the CND may be dragged outward by the central wind. Note that the current set of simulations does not assume central mass loading and the inserted wind source accounts only for the core of the YNC. However, the wind may be mass-loaded outside of the $r_\mathrm{wind}$ region, where the additional mass may originate from either the surface or the inner edge of the CND, from where the gas is exposed to, and ablated by, the wind. 

In summary, the simulation of the gas flow is subjected to the following steps near the GC:
\begin{itemize}
    \item The matter is advected by the \textit{Hydro} (built-in PPM solver) module of \textit{Flash}.
    \item The accreted CND gas reaching the distance $r_\mathrm{GC} < r_\mathrm{sink} = 0.3$ pc is consumed by a sink, if the number density of any cells exceeds density threshold $n_\mathrm{sink} = 500$ cm$^{-3}$. If the gas of a cell within $r_\mathrm{sink}$ is less than this threshold, the accreted material further accumulates until the density reaches $n_{sink}$. 
    %\item If the volume density of the sinked cell is higher than the approximate density the WR stars may ionize ($n_{ion} = 10^4$ cm$^{-3}$) then the matter over this second density threshold is added to the mass of the central wind source, contributing to the internal energy of the wind.
    %\item The sinked matter with density $n_{sink} < n < n_{ion}$ is redistributed as mass loaded wind.
    \item The stellar wind (initially free of any mass-loading) is injected within $r_\mathrm{w0} = 0.5$ pc with a density given by equation (\ref{eq:wind}). 
\end{itemize}

 Since the wind density never exceeds $n_{sink}$, the implementation of the sink within the central wind source does not decrease the mass of the outflowing hot gas. For those cells in which the accretion penetrates to the sink and in which the local density exceeds the $n_{sink}$ limit within $r_{\mathrm{w0}}$), the loss of wind material is negligible.

\begin{figure}
    \centering
	\includegraphics[width=\columnwidth]{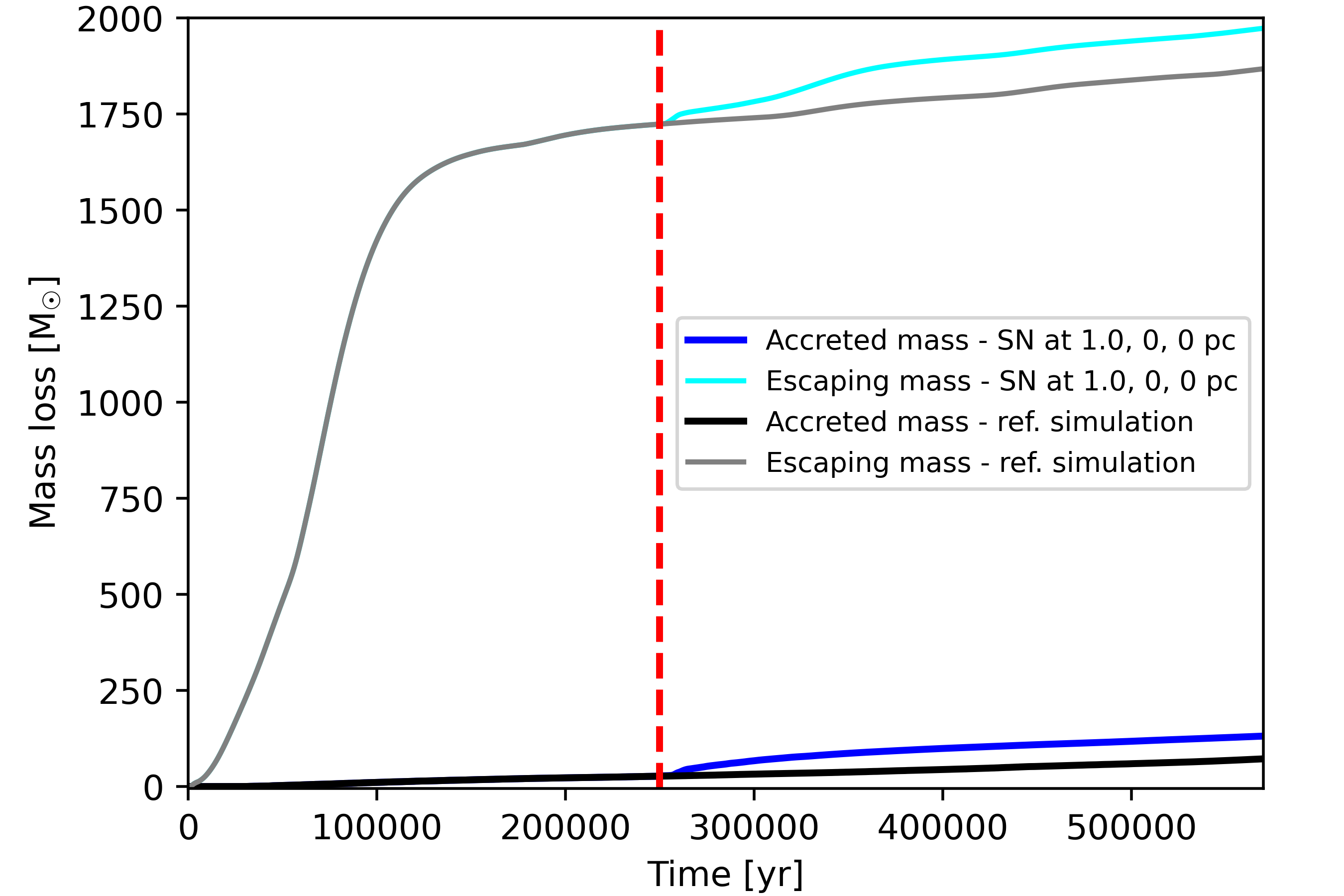}
    \caption{Cumulative mass accretion from the simulation domain with a SN explosion at $t = 250$ kyr (dashed red line indicates this moment). Until the CND relaxes and reaches the MHD-equilibrium (at $\sim$200 kyr), nearly 1700 M$_\odot$ of gas leaves the simulation domain. At the moment of the explosion, the accretion onto the central sink and the escaping mass (in the form of a mass-loaded wind) reach an approximately constant phase. The SN explosion causes an increased rate for both mass flows.}
    \label{fig:sn1_mass_loss}
\end{figure}

\subsection{SN explosions}
\label{sec:explosions}

For the SN simulations, we consistently choose $t = 250$ kyr as the moment of the explosion (the actual status of the disk at this moment can be seen in Fig. \ref{fig:ref_sim1}). Since the Galactic Center is known to host many massive young stars, so that the Type Ia SN rate is significantly lower than that of Type II, we assume that the SN is a typical Type II event with ejecta mass of $M_\mathrm{SN} = 10$ M$_\odot$. The canonical kinetic energy of $E_\mathrm{SN} = 10^{51}$ erg is inserted into the selected region of the grid, in the form of M$_\mathrm{SN}$ of hot gas (although the internal energy is negligible compared to $E_\mathrm{SN}$), modeling the main properties of a typical young core collapse event at the end of the free expansion phase. The $E_\mathrm{SN}$ and $M_\mathrm{SN}$ are inserted on 2x2x2 cells at various explosion sites in the disk to investigate the impact of the SN locations. The SN is expected to affect the dynamics of the CND on a larger scale. Since the kinetic energy is the only attribute of the SN which is comparable to that of the CND, we neglect any other physical properties of the ejecta (e.g., magnetic field, turbulence).

\section{Results}
\label{sec:results}

\subsection{Steady-state accretion}
\label{sec:acc_rate}

The disk reaches MHD quasi-equilibrium in a dynamical timescale ($t_\mathrm{dyn} \simeq 150$ kyr), when the oscillation of physical properties relaxes and both the accretion rate and the rate of mass leaving the domain become constant. Later, the disk keeps its integrity in the reference simulation over at least several dynamical timescales. In this reference model, the accretion of the CND is the result of numerical viscosity, thus, it is a function of the spatial resolution of the simulation, and of the %cogwheel instabilitiy
instabilities due to interaction between stellar winds and CND mass at its inner edge. 

Note that the model accretion is caused by the numerical viscosity which is controlled by the resolution of the simulation.  We aim to perform the simulation with the lowest refinement level which is able to reproduce the accretion rate of the observed CND. For comparison, we follow the description of an $\alpha$-accreting disk \citep{Shakura73} to provide quantitative estimates about the accretion in the CND: 

\begin{equation}
    \dot{M} = -2\pi R \Sigma \mathrm{v}_r ,
\end{equation}
where $\Sigma(R,t)$ is the surface density, and v$_\mathrm{r}$ is the inward radial flow velocity of the disk. 

The $\dot{M}$ is directly measured by the mass accreted by the central sink. Our simulation setup allows us to register the properties of the mass destroyed by the central sink and to distinguish between the mass originating from the CND and the infalling gas from outside of the CND. The contribution of the latter source is less than $1\%$, thus, the mass inflow is almost exclusively accreted from the disk.

The fiducial model of the CND indicates a mass inflow rate of $\dot M_\mathrm{acc} = 1.2 \times 10^{-4}$ M$_\odot$ yr$^{-1}$ over a dynamical timescale, with a constant rate. This value is greater than the Bondi accretion rates estimated for Sgr A*, ranging from $3\times10^{-5}$ to $2\times10^{-4}$ M$_\odot$ \citep[see e.g.][ respectively]{Quataert99, Melia92}, and also exceeds the rate estimated from the observations of the Event Horizon Telescope by orders of magnitude. The difference between observation and the simulation can be interpreted as the majority of accreted mass from the CND gets blown away by the wind from the YNC. Thus, the mass inflow estimated in this study does not represent the direct accretion onto the SMBH, instead the rate at which the matter enters the accretion flow.

The efficiency of the $\alpha$-accreting disk is often described by the $\alpha$ dimensionless scale factor of viscosity:

\begin{equation}
    \alpha = \nu / ( c_s H )
\end{equation}
where $c_\mathrm{S}$ is the local mean (or average over a range of $R$) sound speed, $H$ is the scale height of the disk. The kinematic viscosity, $\nu$ can be estimated as:

\begin{equation}
    \nu \times \Sigma = \frac{\dot{M}}{3 \pi} \Big[1 - (R_* / R)^{1/2}\Big]
    \label{eq:ssacr2}
\end{equation}

where $R_*$ is inner edge of the accretion disk.

The approximate value of $\alpha$ is expected to be $\sim$0.1 for a thin disk around a SMBH. %The contribution due to viscosity may be estimated from equation (\ref{eq:ssacr2}).
It can be expected that $\alpha$ would decrease by increasing the resolution of our model disk, providing a quantitative estimate of the numerical viscosity and a floor value for spatial resolution. In the case of refinement level 6 ($\Delta x = 1.25\times 10^{17}$ cm), $\alpha$ varies between 0.01 and 0.5 depending on the radial coordinate. The external environment of the disk, chosen as outside of $r_\mathrm{GC} = 4$ pc, is resolved with refinement level of 5 ($\Delta x = 2.5\times 10^{17}$ cm).

As the simulations show (see in Fig. \ref{fig:sn1_mass_loss}), the accretion flow in the rotational plane can overcome the ram pressure caused by the wind and reach $r_{sink}$. Outside of $z=0$ (i.e. the rotational plane resolved in two cells width in our simulation), the weaker inward flow of the disk is suppressed by the wind, and no accretion is expected until the addition of shock waves caused by SNe.

\subsection{Impact of supernova explosions}
\label{sec:supernovae)}

We follow the evolution of the CND in seven scenarios assuming a supernova with the same energy exploding at different sites (see in Table \ref{tab:model_data}). For reference, we also evolve the fiducial model setup without any SN event and compare the mass loss history of the disk.  

 The SN properties (described in Sect. \ref{sec:explosions} added to the grid at $t = 250$ kyr quickly produce a shell in the dense environment and after a few tens of kiloyears of evolution, it disperses in the dense CND (an example of the SN shell evolution can be seen in Fig. \ref{fig:sn1_evolution}). Meanwhile, the shock front can reach both the central sink, producing an increment in the accretion, and the low-density outer environment, where the shell quickly leaves the boundaries of the simulation domain. The turbulence caused by the SN also has a long-term effect on the mass transport within the CND. 
 
 The changes in the total mass of the simulation domain can be described as:
\begin{equation}
    \Delta M_\mathrm{total} = \Delta M_\mathrm{wind} - \Delta M_\mathrm{acc} - \Delta M_\mathrm{esc}
\end{equation}
 where the stellar wind mass, $M_\mathrm{wind}$, is assumed to be constant in time. We follow the $M_\mathrm{acc}$ accreted mass, which falls into the central sink (see in Sect. \ref{sec:central}), and the $M_\mathrm{esc}$ mass leaving the simulation domain to characterize the impact of the SN on the CND. The simulations run until both functions reach the same slope as those of the reference model, that is, the point at which no further impact of the explosion can be detected.

The example of the SN\_r10\_i00 run (for the explanation of the nomenclature, see \ref{tab:model_data}), including the first 250 kyr of MHD relaxation phase and the impact of the SN explosion, is shown in Fig. \ref{fig:sn1_mass_loss} compared to the reference model. The extra accretion caused by the inserted kinetic energy of the SN results from the combination of two processes. Right after the explosion, the shock pushes the gas between the center and the explosion site toward the sink causing a rapid growth in accretion at $\sim$50 kyr after an explosion. 
Later, the shock spreading in the disk triggers turbulence, which leads to a loss of  angular momentum of some of the gas, and therefore increased accretion. This latter process is relatively slow and is responsible for less than $\sim$25\% of the extra accretion in SN\_r10\_i00.

\begin{figure}
    \centering
	\includegraphics[width=\columnwidth]{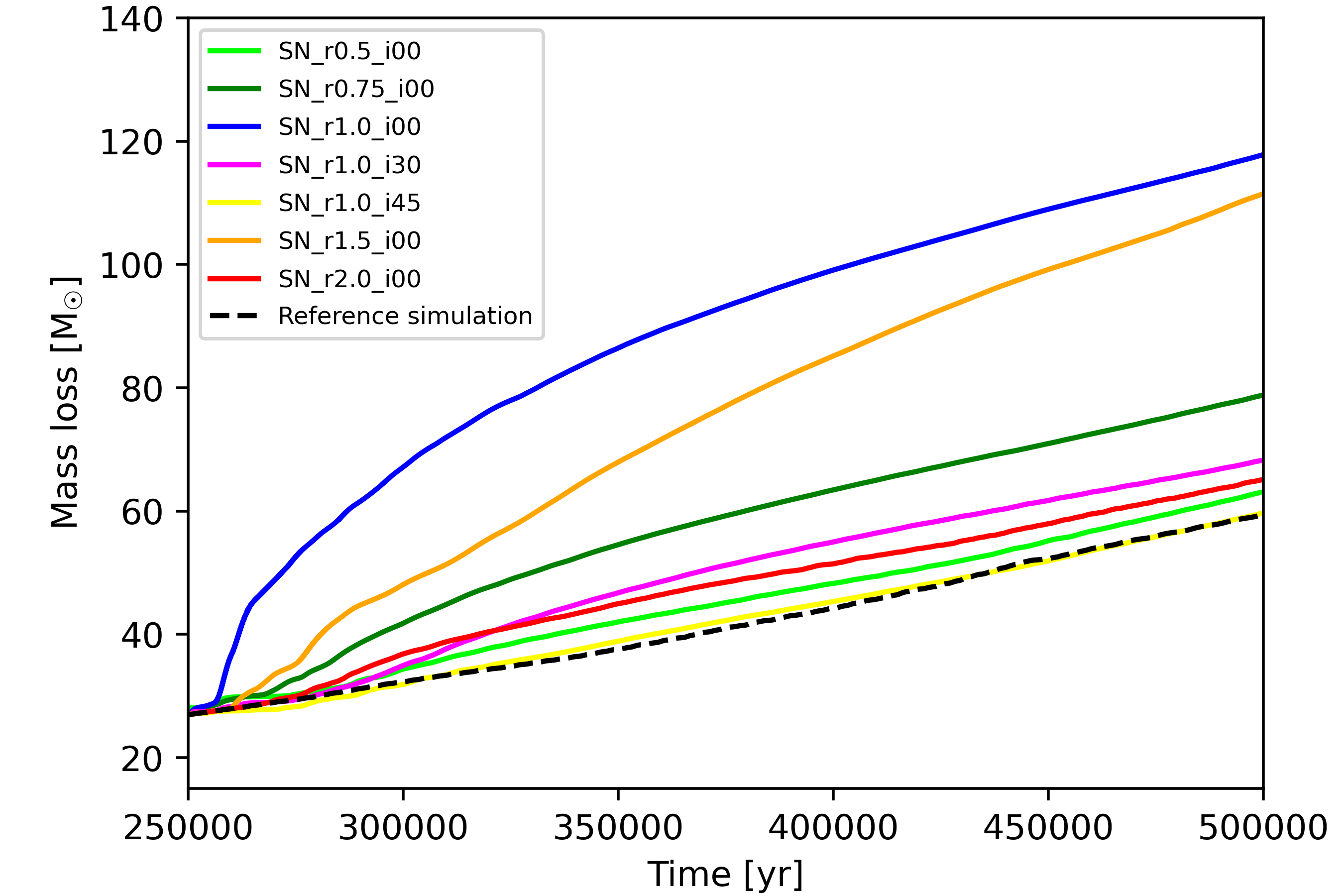}
    \caption{Cumulative accretion into the central sink of CND material after SN events at different explosion sites}
    \label{fig:sn_accretion}
\end{figure}

\begin{figure}
    \centering
	\includegraphics[width=\columnwidth]{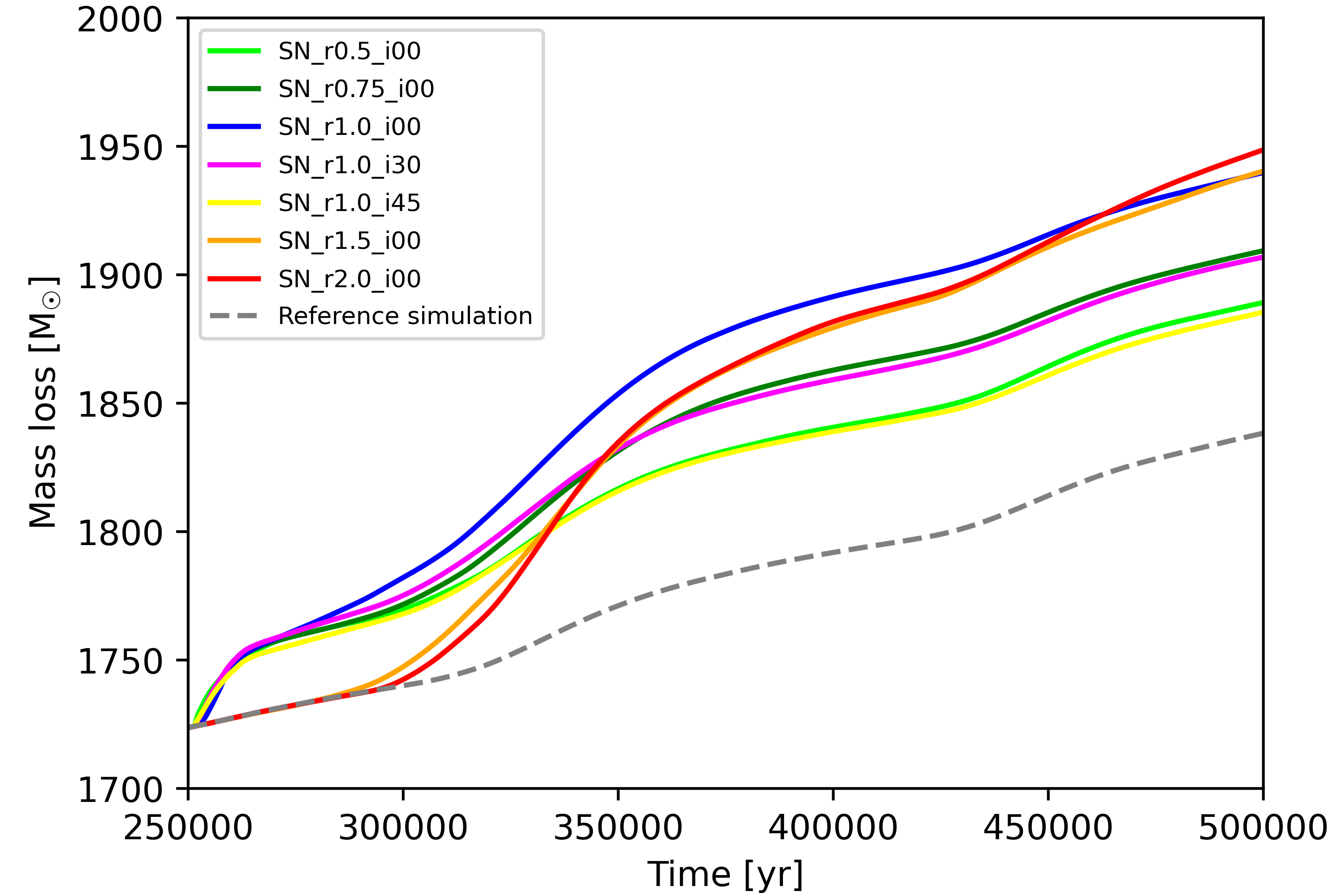}
    \caption{Cumulative mass of gas leaving the simulation domain after SN events at different explosion sites.}
    \label{fig:sn_escaping}
\end{figure}

\begin{figure}
    \centering
	\includegraphics[width=\columnwidth]{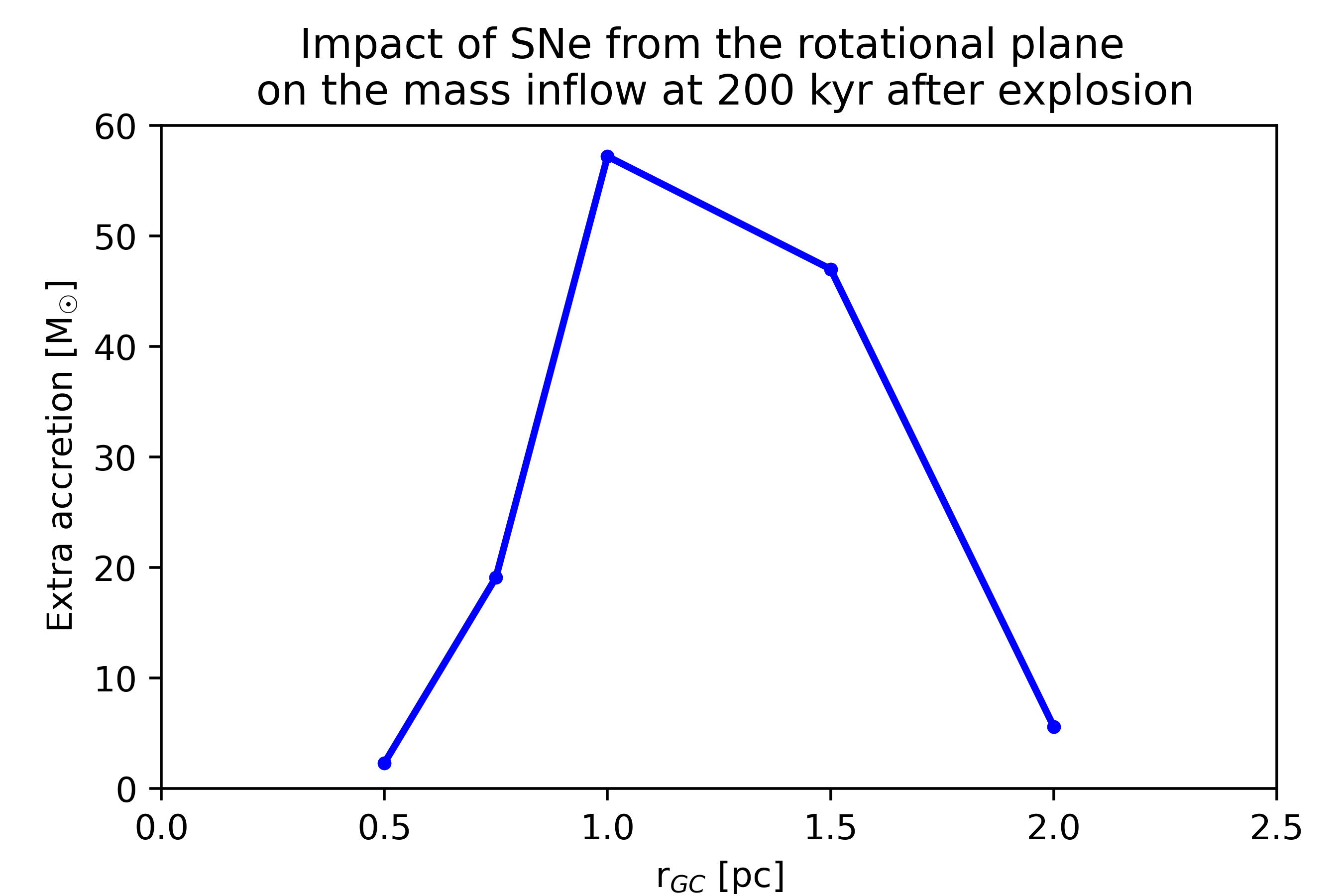}
    \caption{Additional mass inflow caused by SNe occurring in the rotational plane as a function of the galactocentric distance of the explosion site. The accretion is estimated 200 kyr after the explosion ($t=450$ kyr) by subtracting the accretion of the fiducial model.}
    \label{fig:sn_accretion_i00}
\end{figure}

The greatest mass inflow is caused by explosions occurring close to the rotational plane at $r_\mathrm{GC} \simeq 1$ pc (see in Figs. \ref{fig:sn_accretion} and \ref{fig:sn_accretion_i00}). For SNe at closer galactocentric distances (i.e., within the central cavity of the CND), the shock cannot affect the greater part of the disk during its propagation inward, while the kinetic energy of the outer explosions ($r_\mathrm{GC} \geq 1.5$ pc) quickly dissipates in the densest region of the disk. 

The gas leaving the surface of the CND due to kinetic energy insertion quickly expands and leaves the simulation domain. %This process is also supported by the central wind which becomes mass-loaded with a factor of one for a brief period. 
This process is also supported by the central wind which becomes heavily mass-loaded for a brief period (with a factor of $\sim$10 instead of 3, see Sect. \ref{sec:central}). The modeling setup does not allow further following of the blown-out material, but considering that the radial velocity of the SN shock is approximately a few thousand km\,s$^{-1}$, it is likely that the gas can reach high galactic latitudes. The blown-out gas originates mainly from the region having  $r_\mathrm{GC}$ similar to that of the explosion site. The kinetic energy of the SN can effectively sweep out the matter from the lower-density regions of the CND, thus, the relative differences in the cumulative mass of gas escaping the simulation domain are lower compared to those of the accreted material. Depending on the exact explosion site, 50-100 M$_\odot$ leave the vicinity of the CND over 200 kyr (see Fig. \ref{fig:sn_escaping}), after which the growth of the escaping mass stops compared to that of the reference simulation, i.e. the outflow rate returns to normal. Even the most extreme case disturbs $\sim0.3 \%$ of the total mass of the disk. Considering a rough estimate of the SN rate in the studied region of the CND and the continuous mass supply from the inflows of the Central Molecular Zone, the disk will likely remain intact on longer timescales.

   \begin{table}{ccccc}
      \caption[]{Summary of initial SN properties. In each model, the 10 M$_\odot$ ejecta mass is inserted into the grid with 10$^{51}$ erg of kinetic energy. The explosion sites are at r$_{exp}$ galactocentric distance with elevation angle of $\phi$.}
         \label{tab:model_data}
         \begin{tabular}{c c c c c} 
            \hline
            \noalign{\smallskip}
            Model & $r_\mathrm{exp}$ [pc] & $\phi$ [deg]  & $M_\mathrm{ej}$ [$M_\odot$] & $E_\mathrm{kin}$ [erg] \\
            \noalign{\smallskip}
            \hline
            \noalign{\smallskip}
            SN\_r05\_i00 & 0.5 & 0 & 10 & 10$^{51}$   \\
            SN\_r07\_i00 & 0.75 & 0 & 10 & 10$^{51}$  \\
            SN\_r10\_i00 & 1.0 & 0 & 10 & 10$^{51}$\\
	SN\_r10\_i30 & 1.0 & 30 & 10 & 10$^{51}$\\
	SN\_r10\_i45 & 1.0 & 45 & 10 & 10$^{51}$\\
	SN\_r15\_i00 & 1.5 & 0 & 10 & 10$^{51}$\\
	SN\_r20\_i00 & 2.0 & 0 & 10 & 10$^{51}$\\
            \noalign{\smallskip}
            \hline
         \end{tabular} 
   \end{table}

\section{Conclusions}
\label{sec:conclusions}

We have conducted 3D magneto-hydrodynamic simulations of the long-term evolution of the Galactic CND and have evaluated the impact of SNe exploding within it or in its vicinity.

The simulations consist of the initially warm CND with mass $M_\mathrm{CND}=22000$ M$_\odot$ in a torus-like disk and the wind of $\sim$100 massive stars of the young nuclear star cluster.  Apart from magneto-hydrodynamics, the simulations take into account the gravitational potential of the SMBH and the NSC, and radiative cooling of the gas. After the simulation reaches MHD quasi-equilibrium, we test the impact of SN explosions at various locations in the CND.

\begin{itemize}
    \item The shock wave of the SN transports significantly more mass out of the disk than the original ejecta mass, over less than a dynamical timescale. The blown-away gas is faster than the escape velocity at the edge of the simulation domain, thus it is likely to reach high Galactic latitudes. The amount of escaping mass per SN explosion is between $\sim$50 M$_\odot$ (for explosion sites closer to the edge of the disk) and $\sim$100 M$_\odot$ (for explosion sites in the rotational plane).
    \item After the SN explosion, all simulations showed additional mass inflow to the central sink compared to the steady-state accretion of the fiducial model without any SNe. The additional mass transport is between 2 and 60 M$_\odot$ depending on the location of the explosion site. The greatest impacts on the accretion rate are caused by SNe exploding in the rotational plane at distances of 1 - 1.5 pc from the GC.
    \item Although the estimated mass inflows triggered by the SN explosions are significantly lower than those of simulations assuming a homogeneous ambient  \citep{Palous20, Barna22}, our models demonstrate that gas swept up by SN shocks can overcome the ram pressure of the central wind source and provoke increased accretion toward the close vicinity of Sgr A*. Note that our simulations cover only a handful of selected explosion sites, but a larger sample of SN locations and a more established SN rate an estimate of the frequency of enhanced accretion events, as well as the enhanced mass loss rate during those events.  
\end{itemize}

This analysis is restricted to the GC of the Milky Way as the specific properties of the environment (e.g. gravitational potential, central wind, mass of the CND) have strong impacts on the outcome of the simulation. Modeling the centers of other galaxies, which will be the subject of future studies, is necessary to understand the contributions of supernovae to the  growth rates and activities of SMBHs elsewhere.

\begin{acknowledgements}
  BB received support from the Hungarian National Research, Development and Innovation Office grants OTKA PD-147091.  R.W., J.P., and S.E. acknowledge support by the project RVO:67985815. P. V. was supported by French government through the National Research Agency (ANR) with funding grant ANR AGN\_MELBa (ANR-21-CE31-0011)
\end{acknowledgements}

\bibliographystyle{aa}
\bibliography{aanda}

\end{document}